\documentclass[
 amsmath,amssymb,
 reprint,%
]{revtex4-1}

\usepackage{graphicx}% Include figure files
\usepackage{dcolumn}% Align table columns on decimal point
\usepackage{bm}% bold math
\usepackage[usenames,dvipsnames]{color} %using the color package, not xcolor
\usepackage[utf8]{inputenc}
\usepackage[T1]{fontenc}
\usepackage{mathptmx}
\usepackage{amsmath}
\usepackage{etoolbox}
\usepackage{ulem}
\usepackage{xcolor}
\usepackage{natbib}
\makeatletter
\def\@email#1#2{%
 \endgroup
 \patchcmd{\titleblock@produce}
  {\frontmatter@RRAPformat}
  {\frontmatter@RRAPformat{\produce@RRAP{*#1\href{mailto:#2}{#2}}}\frontmatter@RRAPformat}%
  {}{}
}%
\makeatother
\DeclareUnicodeCharacter{2212}{-}
\begin{document}
\preprint{AIP/123-QED}

\title[Sample title]{Post disruption reconnection event driven by a runaway current}
% Force line breaks with \\
\author{L. Singh$^{1,2}$} 
\author{D. Borgogno$^1$}%
\author{F. Subba$^2$}%
\author{D. Grasso$^1$}%

\affiliation{$^1$
Istituto dei Sistemi Complessi—CNR and Dipartimento di Energia, Politecnico di Torino,
Torino, Italy}
\affiliation{$^2$
NEMO Group, Dipartimento Energia, Politecnico di Torino, Torino, Italy}

\date{\today}% It is always \today, today,

\begin{abstract}
The role of a runaway current in a post disruption plasma is investigated through numerical simulations in an asymmetric magnetic reconnection event. While the runaways do not alter the linear growth of the island, they lead to a rotation of the island in the poloidal direction as found in  [C. Liu {et al.} Physics of Plasmas 27, 092507 (2020)]. The role of a microlayer smaller than the resistive one is thoroughly investigated. While the resistive layer controls the transition of the island from the linear to the nonlinear stage, the microlayer width causes the runaways to become nonlinear as soon as the size of the island exceeds it. Moreover, this transition of the runways electrons to the nonlinear phase is accompanied by a drastic redistribution of runaways within the island with respect to the symmetric case. The influence of the electron skin depth on the linear evolution is also taken into account. Finally, nonlinear simulations show that the rotation frequency tends toward zero when the island saturates.
\end{abstract}

\maketitle

\section{\label{sec:level0}Introduction}

Runaway electrons are a major cause of concern for future fusion devices, ITER \cite{Boozer15} foremost among them.  Because of the decrease in electron collision frequency with increasing velocity, electrons subjected to a strong electric field can experience unlimited "runaway" acceleration. In tokamaks, runaway electrons can be produced in the disruptions because of the strong inductive electric field formed when the thermal energy of the plasma is rapidly lost. The runaway population can grow exponentially (avalanche mechanism) due to collisions of relativistic electrons with low-energy electrons.\\
It is estimated that the impact of electron runaways on the walls of a machine such as ITER can cause severe damage. It is therefore important to improve the understanding of plasma dynamics during all the life time of the RE beam to mitigate its effects. Recently \cite{Bandaru2021} a study has been carried out to model the RE benign termination observed in the JET shot 95135 \cite{Reux2021}. In particular, during this shot, MHD activity has caused the mitigation of RE through a broad deposition of the beam on plasma-facing components. It has been demonstrated in  \cite{Bandaru2021}, through numerical simulations done with JOREK \cite{Hoelzl2021}, that this RE suppression is caused by the stochastization of the magnetic field lines causing the crash of the RE current. \\ 
In this perspective, it also becomes important to understand what is the interaction mechanism between the runaway population generated during a disruption and the post-disruption plasma. That is, to understand what are the stability properties of such a plasma in which the plasma current is replaced by the runaway current. More than a decade ago this problem was studied in \cite{Helander2007} by considering the role of a runaway electron current on the spontaneous development of the magnetic reconnection instability in a resistive, weakly unstable plasma. 
It was found that when the plasma current is totally carried by runaway electrons, two-dimensional (2D) magnetic perturbations, with no dependence on the spatial coordinate along the guiding magnetic field, which resonate on the surface where the current peaks, significantly increase the saturated amplitude of the reconnected region (magnetic island) compared to the standard case without runaway electrons, but do not affect the linear growth rates. More recently \cite{Grasso2022} some aspects, which had remained unclear in \cite{Helander2007} because of the periodic equilibrium magnetic configuration adopted, were clarified with new numerical simulations assuming a different equilibrium. Indeed, the periodic equilibrium configuration implied the presence of two magnetic islands, one at the center of the integration domain and one at the boundaries. When the magnetic islands grew too large in the nonlinear phase they began to interact, affecting the saturation results.\\
In 2020, Liu et al. \cite{Liu2020} extended this analysis by considering the linear evolution of 2D asymmetric perturbations, whose resonant surfaces do not lie at the current peaks. Two peculiar features have been highlighted: firstly, the poloidal rotation of the magnetic island, the frequency of which depends on the derivative of the runaway electron current on the resonant surface, and secondly, the existence of a microscopic layer in which the current density of runaway electrons is concentrated. Since the width of these layers depends on the inverse of the runaway electron velocity, that is of the order of the speed of light, it is expected that they can thin to microscales where the kinetic effects can make a non-negligible contribution to plasma dynamics.  
This paper is focused on the selfconsistent study of the mutual interaction between runaway electrons and asymmetric magnetic reconnection in nonlinear regimes. The analysis is performed by considering a two-fluid, collisional plasma model, where the effects of the electron inertia and the electron temperature, that introduce the electron skin depth and the ion-sound Larmor radius, respectively, are taken into account. Although these scales are small in a post disruption scenario due to the low plasma temperature and the electron mass, they could become relevant in presence of localized current layers, affecting the evolution of the global process. The linear analysis of a RE-driven magnetic reconnection reproduced the results obtained in \cite{Helander2007}  with RE not having an influence on the linear growth rates of the island both in symmetric and asymmetric cases and the island width at saturation being $50\%$ higher with RE with respect to the case without RE. With respect to the symmetric case, an asymmetric current profile leads to the island rotation and the presence of a microlayer on the RE current distribution at the X-point consistently with the results shown in \cite{Liu2020}. In addition, the electron skin depth affects the thermal electron distribution at the X-point of the island. The nonlinear evolution of asymmetric modes leads to the generation of a spiral-like structure inside the island whereas the island rotation frequency tends towards zero locking in at saturation.  \\
The paper is organized as follows. In section \ref{sec:level1} the model equations are introduced. In section \ref{sec:level2} the SCOPE3D  numerical tool is presented and its benchmark and verification test is shown in \ref{sec:level3}. Sections \ref{sec:level4} focus on linear and nonlinear results for asymmetric modes rispectivley.  Conclusions close the paper.

\section{\label{sec:level1}Model equations}

In our analysis we extend the reduced, purely collisional model adopted in \cite{Helander2007} by considering as in ref. \cite{Grasso2022}  the contributions of the effects of the electron mass $m_e$ and the electron temperature $T_e$  through new terms in the plasma Ohm's law proportional to the electron skin depth $d_e=c/\sqrt{4\pi n_e e^2/m_e} $ and the ion sound Larmor radius $\rho_s=\sqrt{(T_e/m_i)}/\omega_{ci}$, with $\omega_{ci}$ the ion gyrofrequency, respectively. Furthermore, here we consider a three-dimensional slab geometry, which also allows us to deal with perturbations dependent on the coordinate along the direction of the guiding magnetic field direction. 
%In Liu et al. [6] the existence of a sublayer was found inside the resistive layer that leads to a real frequency in the magnetic island growth. Since this sublayer  has a width comparable to the electron skin depth, we do not exclude the contribution of electron inertia in the plasma Ohm's law even though a post disruption plasma is often characterized by high collisionality.
The equations normalized on the Alfven time and on the characteristic length of variation of the equilibrium magnetic field are \cite{Grasso2022}:
\begin{eqnarray}
\frac{\partial \psi}{\partial t} + [\varphi,\psi] + d_e^2 \frac{\partial J}{\partial t} + d_e^2[\varphi,J] -\rho_s^2[U,\psi]\nonumber\\
 +\eta (J-J_{RE}) +\frac{\partial \varphi}{\partial z} 
 +\rho_s^2\frac{\partial U}{\partial z}= 0 \label{psieq}
\end{eqnarray}
\begin{eqnarray}
\frac{\partial U}{\partial t} + [\varphi,U] - [J,\psi] -\frac{\partial J}{\partial z}= 0
\label{Ueq}
\end{eqnarray}
\begin{eqnarray}
\frac{\partial J_{RE}}{\partial t} + [\varphi,J_{RE}]+\frac{c}{v_A}([\psi,J_{RE}]-\frac{\partial J_{RE}}{\partial z})  = 0
\label{Jre}
\end{eqnarray}
\begin{eqnarray}
J = - {\nabla^2}_\perp\psi,   U = {\nabla^2}_\perp\varphi  
\label{closure}
\end{eqnarray}
where $ {\nabla^2}_\perp = \partial^2_x + \partial^2_y$ and $[f, g]=\partial_x f \partial_y g -\partial_x g\partial_y f$. The model assumes a magnetic field $B = B_0\bm{e_z} + \nabla\psi \times \bm{e_z}$, where $B_0$ represents the uniform magnetic guide field and is set to $1$, and a velocity field $v = -\nabla\varphi \times \bm{e_z}$. The fields $\psi$ and $\varphi$ are the magnetic flux and the stream function, respectively, while $J$ is the current density of the plasma and $U$ its vorticity. Eq. \ref{Jre} describes the evolution of the runaway current density $J_{RE}$, where it has been made the assumption that these particles move with the (normalized) speed of light along the magnetic field lines. Due to their relativistic velocity, in contrast to the thermal electrons, runaway electrons do not collide with ions, as stated by the dissipative term in the plasma Ohm's law in Eq. \ref{psieq}, where $\eta$ is the normalized resistivity.  
%\begin{eqnarray}
%\frac{\partial J_{RE}}{\partial t} + [\varphi,J_{RE}]+\frac{c}{v_A}([\psi,J_{RE}]-\frac{\partial J_{RE}}%{\partial z})  = 0
%\label{Jre}
%\end{eqnarray}
%where the runaways are assumed to travel at the speed of light.

In this paper, we focus on the analysis of spontaneous magnetic reconnection events induced by single helicity (SH) perturbations in a sheared, unstable equilibrium magnetic configuration, with $B_y = B_y(x)$. For a generic field, $f$, SH modes have the following  form:
\begin{equation}
f(x,y,0)=\sum_{k_y,k_z} \hat{f}_{k_y,k_z}(x)\exp(ik_yy +ik_zz)\nonumber\\
=\sum_{k_y} \hat{f}_{k_y}(x)\exp(ik_y(y +k_z/k_yz))
\end{equation}
where the helicity $\alpha=k_z/k_y$ is fixed. Here $k_y = \pi m/L_y$ and $k_z = \pi n/L_z$, with $m,n$  integer numbers and $Ly$ and $L_z$ the half-widths of the computational box along $y$ and $z$, respectively. As shown in \cite{Grasso2020, Borgogno2005}, SH problems can be treated as 2D problems, i.e. without dependence on the $z$ coordinate, by transforming the sheared component of the equilibrium magnetic field from $B_y$ to $B_y - \alpha$, which corresponds to a rotation in the $(y,z)$ plane.

\section{\label{sec:level2}The numerical tool SCOPE3D}
The SCOPE3D (Solver for COllisionless Plasma Equations in a 3D slab geometry) code is adopted to solve the equations \ref{psieq}-\ref{closure} in a slab geometry. It is based on an explicit, third-order Adam-Bashforth temporal discretization and is parallelized along the two periodic directions $y$ and $z$. In order to have a high spatial resolution in the reconnection region, a compact finite difference scheme \cite{Lele1992}, specifically designed for a non-equispaced grid, is adopted for the spatial discretization along the x direction.  Fast Fourier methods are applied instead along the periodic directions. In addition, numerical filters are used in the y and z directions to remove short-length scales caused by nonlinear interactions \cite{Lele1992}, while  the physical dissipation is sufficient to control the numerical noise along the $x$ direction. Since we analyze here only SH modes  all the simulations have been carried out in the 2D limit, saving computational time.\\
We consider magnetic reconnection events starting from a static plasma, immersed in an asymmetric, Harris-type, sheared magnetic field \cite{Haris1962}, in which all current is carried by runaway electrons, such that:
\begin{equation}
%\varphi(x,y,0)=0
\varphi_{eq}(x)=0
\end{equation}
\begin{equation}
%\psi(x,y,0)=\psi_{eq}(x) + \hat{\psi}(x)\exp(ik_yy) +c.c.
\psi_{eq}(x)=-\log(\cosh(x))+\alpha x
\label{Harris}
\end{equation}
\begin{equation}
 J_{RE_{eq}}(x)=J_{eq}(x)=-\nabla^2\psi_{eq}(x)
\end{equation}
Concerning the grid used in this work, for the linear analysis a resolution of ny=96 points has been used along the periodic direction y. In contrast, for the x-direction,  the number of grid points has been varied from 1200 to 4800 on the nonequispaced grid in order to have an adequate resolution in the reconnection region. In particular, for the purpose of benchmarking, nx = 1200 grid points have been adopted for the x-direction so as to guarantee a resolution of dx = 0.0038 around x=0 where the reconnection occurs.

\section{\label{sec:level3}  SYMMETRIC CASE: $\alpha$=0}
\begin{figure}
\includegraphics[width=9cm, height=6cm]{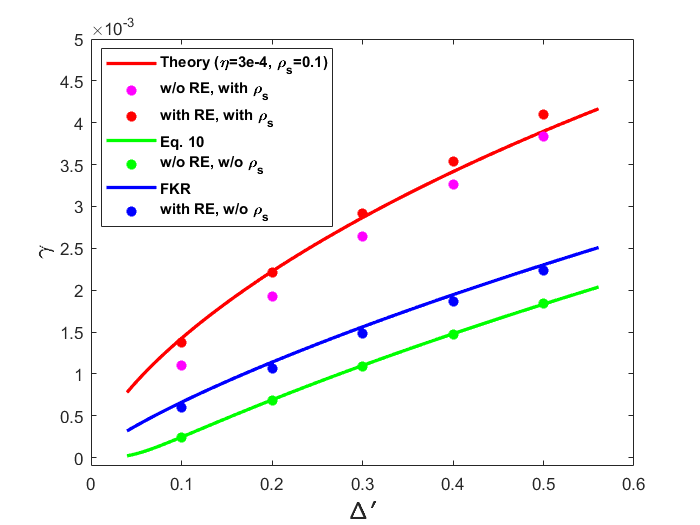}% 
\caption{\label{fig:grrun} Numerical and analytically derived linear growth rates for a pure resistive reconnecting mode driven by a runaway current (blue points and blue curve) compared with the case without runaway current (green points and green curve) along with the numerical linear growth rates with (red points) and without a runaway current (magenta points) compared with the results of Eq.  \ref{growth_runaways} in presence of electron temperature effects (red curve).}
\end{figure}

Here we briefly summarize the linear and nonlinear results \cite{Grasso2022} we obtained assuming a runaway current profile peaked on the rational surface at $x=0$,  in order to have at hands a  comparison term when exploring the $\alpha \neq 0$ cases. These results have been validated against the ones reported in \cite{Helander2007} and  for this reason we consider the pure resistive regime ($d_e = 0$) and the limit $c/v_A = 1$.  A $[-L_x, L_x]\times[-L_y, L_y]$ with $L_x = 3\pi$ and different $L_y$ domain was used to integrate the equations. $L_y$ was varied to account for different degrees of instability. \\
Differently from \cite{Helander2007}, where a periodic in-plane component of the equilibrium magnetic field was adopted, i.e. $\psi_{eq}=\cos(x)$, the equilibrium (\ref{Harris}) allowed us to carry out analysis on long nonlinear times and to investigate the evolution of a single magnetic island until saturation. This has not been possible in \cite{Helander2007} since the periodic equilibrium adopted there led to the presence of a second magnetic island at the boundaries of the integration domain, influencing the one located at the center. The Harris equilibrium leads to an equilibrium current that has been discussed in MHD theory as the most probable profile \cite{Biskamp2000}.
Figure \ref{fig:grrun} shows the numerical and analytical linear growth rates with and without runaways. The resistivity  is fixed at $\eta = 3e- 4$ and different perturbation wave numbers $k_y$ are adopted, corresponding to modes with different values of the stability parameter $\Delta'=2(1/k_y - k_y)$. Furthermore, the effect of electron compressibility along magnetic field lines is taken into account through the introduction of the ion sound Larmor radius scale length,  $\rho_s$, into the equations. In particular, the blue points for runaways and green points for no runaways  are compared with the expected theoretical values with (blue curve) and without (green curve) runaways. The theoretical prediction given by Eq. \ref{grnorun}  and represented by the green curve, accounts for the finite resistivity correction because of the relatively high value of $\eta$ \cite{Militello2004}. In the same  figure red and magenta  points  represent the linear growth rates in presence of runaways and  when no runaways are present fore cases where $\rho_s=0.1$. These points are compared with the results of Eq. \ref{growth_runaways}, corresponding to the red curve. It can be observed that the runaway current's presence does not significantly alter the linear growth rates as in the pure resistive case.
The linear dispersion relation shown in fig. \ref{fig:grrun} has been derived in slab geometry in Ref. [5] with and without RE, obtaining:

\begin{equation}
\frac{\gamma^{5/4}}{\eta^{3/4}k_y^{1/2}}=0.47\Delta' ~~~~~~~~      \text{with runaways}  \label{grrun}
\end{equation}
\begin{equation}
\frac{\gamma^{1/4}(\gamma-2b\eta)}{\eta^{3/4}k_y^{1/2}}=0.47\Delta' ~~~~~~~~ \text{without runaways}
\label{grnorun}
\end{equation}
\noindent where $b = \frac{\psi_{eq}^{IV}}{\psi_{eq}^{II}}\mid_{x=0}$.

Eq. \ref{grrun} is the standard Furth, Killeen and Rosenbluth (FKR) \cite{FKR1963} growth rate in the small $\Delta'$ regime, defined by the inequality $\Delta'\eta^{1/3}\ll1$. While the derivation of the eq. \ref{grnorun} considers the higher-order derivatives corrections of the current density at the resonant surface \cite{Militello2004}. On the other hand, in presence of $\rho_s$ the dispersion relation for the linear growth rates becomes \cite{Porcelli91},
\begin{equation}
    \frac{\gamma^{3/2}}{\eta^{1/2}k_y}\gamma = 0.32\rho_s\Delta'    \label{growth_runaways} 
\end{equation} 
As it can be observed in fig. \ref{fig:grrun}, a good agreement is found in both cases.
In fig. \ref{fig:island} we compare the nonlinear saturated magnetic island widths in presence  and absence of runaways, $w$, for $\Delta'$ in the range [0.1, 2], where $w$ is given by \cite{Helander2007}:
\begin{figure}[h]
\includegraphics[width=9 cm, height=6cm]{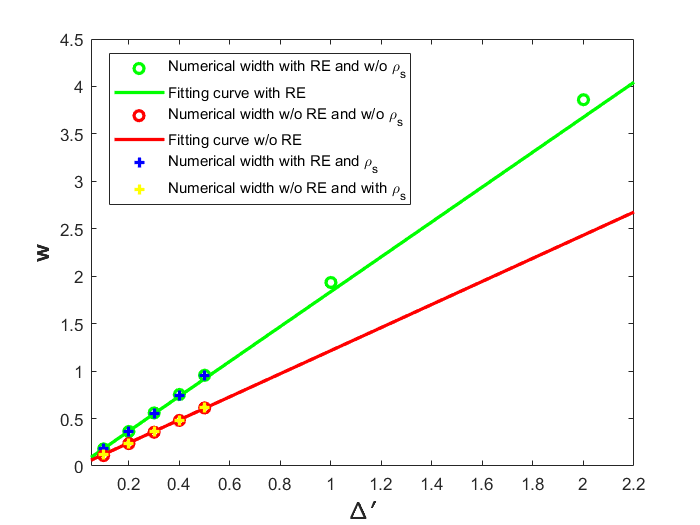}% 
\caption{\label{fig:island} Numerical and analytically derived saturation island widths for the pure resistive reconnecting mode driven by a runaway current (green points and green line) compared with the case without runaways (red points and red line) along with the numerical saturated island widths with (blue plus markers) and without (yellow plus markers) in presence of electron temperature effects.}
\end{figure}

\begin{equation}
w=-\frac{1}{b}\frac{\Delta'}{0.272} ~~~~~~~~ \text{with runaways}  
\label{wrun}
\end{equation}
\begin{equation}
w=-\frac{1}{b}\frac{\Delta'}{0.411} ~~~~~~~~ \text{without runaways}.
\label{wnorun}
\end{equation}

Having $b = -2$ for the type of equilibrium considered in this study, we get $w = 1.85\Delta'$ and $w = 1.22\Delta'$ respectively for the case with and without RE. 
As found in  ref. \cite{Helander2007}, the presence of a runaway current leads to an increase of $50\%$ in the saturated magnetic island width with respect to the case with no RE.\\ 
The analytical expressions found for the saturated island widths in presence, represented by the green line, and absence, represented by the red line, of runaways are plotted in fig. \ref{fig:island} to compare them against the numerically obtained results shown with the corresponding color points. In addition,  fig. \ref{fig:island} shows the numerically obtained saturated magnetic island widths in presence of $\rho_s$ with and without runaways corresponding to blue and yellow plus markers respectively. As it can be observed, the presence of the electron temperature does not lead to any change in the saturation width since the microphysics at the ion sound Larmor radius scale plays no role in the saturated island width, which should depend only on the free energy available for the reconnection process.\\
A good agreement is found between the theory and simulation, however, a difference between the theoretical and the numerically observed widths in presence of runaways can be observed at values of $\Delta'$ of order unity. This follows from the fact that at these $\Delta'$ values, the simulation parameters do not fall into the range of validity of the analytical theory since they depart from the asymptotic limit of the small $\Delta'$ regime. However, these simulations were necessary in order to verify the presence of a bifurcation in the sequence of saturated equilibria which was postulated in ref. \cite{Helander2007}. In our study we did not find any bifurcation allowing the island to grow up to saturation even for values of $\Delta'$ of order 1. In ref. \cite{Helander2007} periodic boundary conditions in the inhomogeneity direction of the equilibrium magnetic field prevented the island to reach a saturated state.\\

\section{\label{sec:level4} ASYMMETRIC CASE $k_z/k_y\neq0$}
Here we consider asymmetric modes, by shifting the  in-plane component of the equilibrium magnetic field $B_{y_{eq}} = \text{tanh}(x) -\alpha$. Hence the rational surface is now located at $x_s=\mbox{settanh}(\alpha)$  while the runaway current profile is peaked at $x=0$.

\subsection{\label{asymlin} LINEAR ANALYSIS} 
When considering  modes for which  $k_z/k_y\neq0$ Liu et al. \cite{Liu2020}  demonstrate that the  runaway electrons convection causes a mode rotation which explains the real frequency seen in their simulation campaign carried out  with the code M3DC-1. In the same work, RE are shown to lead to the formation of a smaller layer within the resistive layer, which width depends on the ratio $c/v_A$. In particular, the resistive layer half width, $\delta_1$ (referred as the layer width in the following), the sublayer half width, $\delta_2$ (referred as the sublayer width in the following), and the growth rate  are given in slab geometry by  
%\begin{equation}
%\delta_2=\gamma v_A/{k_yc}
 %   \label{delta2_Liu}
%\end{equation}
%In this case, the expression for the growth rate becomes
%\begin{equation}
 %   \frac{\gamma^{5/4}}{\eta^{3/4}k_c{1/2}} \frac{2\pi\Gamma(3/4)}{\Gamma(1/4)}=\Delta'-i\pi\frac{mJ'_{RE0}}{|k_c|r_s}
 %   \label{Gr_Liu}
%\end{equation}
%With respect to  Liu et al. \colorbox{BurntOrange}{Mettere il numero della referenza}, in our analysis some changes were implemented  to make the expressions compatible with a 2D slab geometry configuration. Therefore, 
\begin{equation}
   \delta_1=\gamma^{1/4}\eta^{1/4}k_y^{-1/2} 
   \label{delta_1}
\end{equation}
\begin{equation}
    \delta_2=\gamma v_A/{k_yc}
    \label{delta_2}
\end{equation}
\begin{equation}
    \frac{\gamma^{5/4}}{\eta^{3/4}k_y^{1/2}} \frac{2\pi\Gamma(3/4)}{\Gamma(1/4)}=\Delta'-i\pi\frac{k_yJ'_{RE0}}{|k_y|}
    \label{Gr}
\end{equation}
where $J'_{RE0}$ is the derivative of the RE current at the rational surface. For $J'_{RE0}\neq0$ an imaginary part of the growth rate appears which gives a rotation of the magnetic island.
%introduced in Equation \ref{Gr_Liu} to account for the effect of RE on the resistive tearing mode.\\
%\textcolor{red}{To perturb a single helicity mode   in the numerical campaign} the initial condition for $\psi$ is modified according to ATTENZIONE: NOI ABBIAMO CAMBIATO L'EQULIBRIO NON LA PERTURBAZIONE:
%\begin{equation}
%\psi(x,y,0)=\psi_eq(x) + \hat{\psi}(x)\exp(ik_yy +ik_zz) +c.c.
%\end{equation}
%where $k_z = \pi n/L_z$, with n an integer number, is the wave vector along the z-direction.\\ 
The $\Delta'$ parameter taking into account corrections due to finite values of  $k_z/k_y$ has been evaluated according to \cite{Daughton2011},
\begin{equation}
\Delta'\approx 2\left(\frac{1}{k}-k\right)\left[ 1+\frac{\text{tanh}^2{x_s}}{2}\left(1+\frac{1}{1-k}\right) \right]
\label{Daughton}
\end{equation}
where $k\equiv |\textbf{k}|$ with 
$\textbf{k}= k_y\bm{e_y} + k_z\bm{e_z} $.
%The Eq. \ref{Daughton} allows computing correctly the $\Delta'$ parameter when oblique tearing modes are considered. 

\begin{figure}[h]
\includegraphics[width=9 cm, height=6cm]{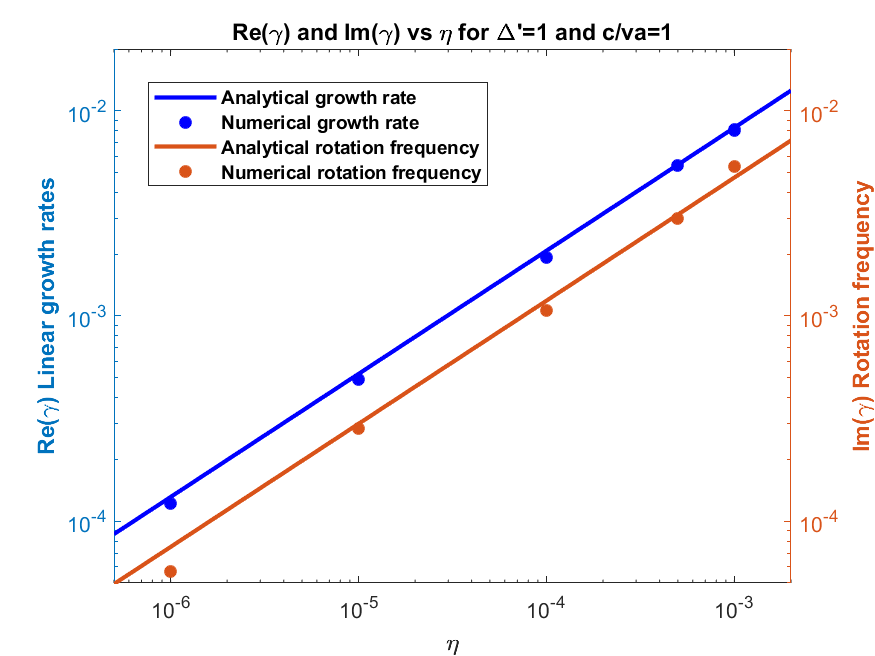}% 
\caption{\label{fig:gammaomega_1_cv1}Comparison between the analytical growth rate (blue line) and rotation frequency (rust curve) given by Eq. \ref{Gr} and the numerical growth rate (blue points) and rotation frequency (rust points) for values  of $\eta$ in the [1e-3, 1e-6] range, $\Delta'=1.015$ and $c/v_A=1$.}
\end{figure}

In fig. \ref{fig:gammaomega_1_cv1} the growth rates, given by the Eq. \ref{Gr}, are shown by the blue line and compared with the simulation results (blue points) for cases with $\Delta'=1.015$, $c/v_A=1$ and resistivity values in the [1e-3, 1e-6] range. As can be observed, the simulation results agree well with the analytical derivation and the same is true for the rotation frequency represented by the rust points and compared  with the results of the Eq. \ref{Gr} (rust curve). 

\begin{figure}[h]
\includegraphics[width=9 cm, height=6cm]{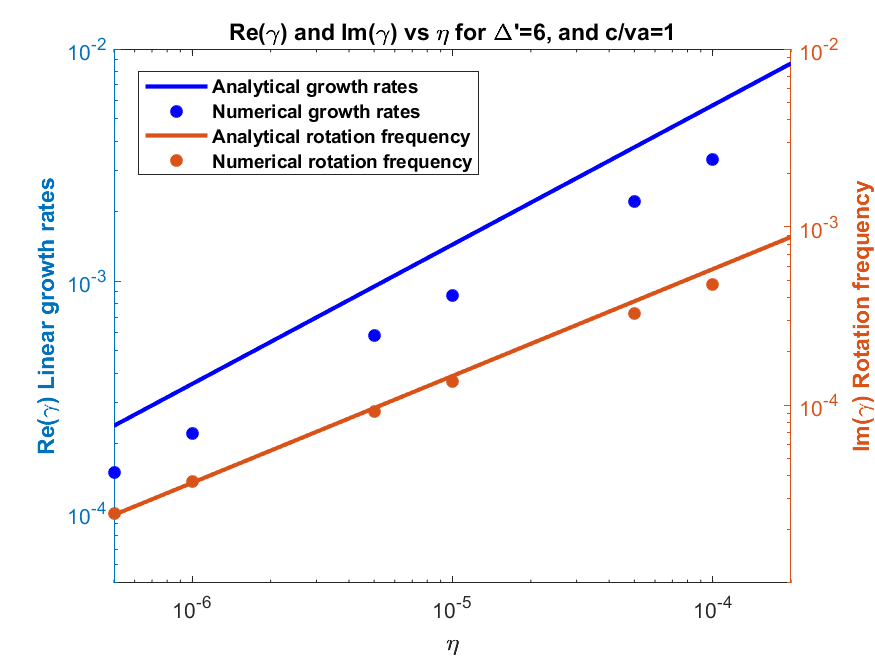}% 
\caption{\label{fig:gammaomega_6_cv1}Comparison between the analytical growth rate (blue line) and rotation frequency (rust curve) given by Eq. \ref{Gr} and the numerical growth rate (blue points) and rotation frequency (rust points) for values  of $\eta$ in the [1e-4, 5e-7] range, $\Delta'=6.057$ and $c/v_A=1$.}
\end{figure}

To analyze the inner layer, we have chosen the value $\Delta'=6.057$, even though fig. \ref{fig:gammaomega_6_cv1} shows that the growth rates from the simulations (blue points) are less close to the analytical results (blue curve), than the more asymptotic case $\Delta'=1.051$. At the same time, the resistivity range considered for this case has been changed to [1e-4, 5e-7] with respect to the range adopted for the previous case in order to respect the small $\Delta'$ regime limit.
%since the small $\Delta'$ regime limit $\Delta'\eta^{1/3}\ll1$ is  more observed for the latter $\Delta'$. 
On the other hand, the $\Delta'=1.051$ case does not allow a sufficient  numerical resolution to determine the inner layer widths already with $c/v_A=1$, so that analysis with $c/v_A=10$ is computationally infeasible. An accurate measurement of the $\delta_2$ requires adopting an increasing number of grid points along the radial direction with decreasing resistivity.  This leads to reducing the radial extension of the simulation domain in order to have enough spatial resolution in the reconnection region. As a consequence, with a smaller domain, the system is less asymptotic which partially explains also the differences in the observed growth rates in fig. \ref{fig:gammaomega_6_cv1} Furthermore, the numerically obtained rotation frequencies (rust points) for $\Delta'=6.057$ agree with the theory (rust curve), as can be seen in fig. \ref{fig:gammaomega_6_cv1}, since the rotation frequency does not depend on the value of $\Delta'$, in accordance with the Eq. \ref{Gr}.

\begin{figure}[h]
\includegraphics[width=9 cm, height=6cm]{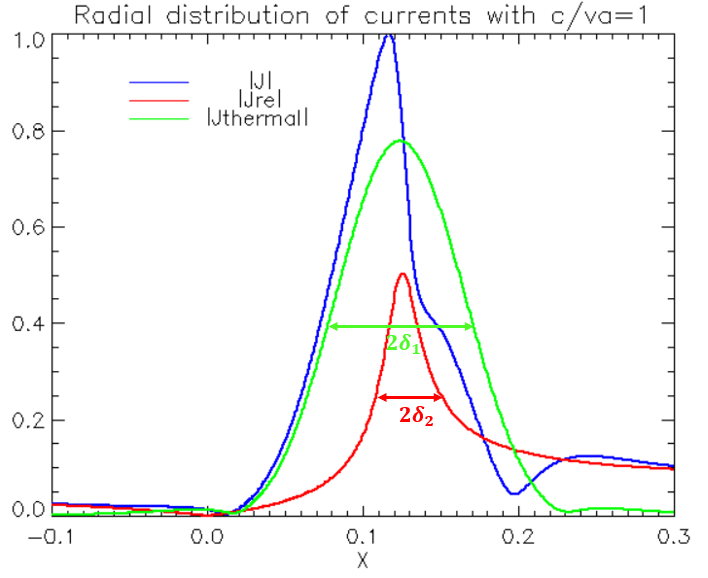}% 
\caption{\label{fig:Delta2_delta1_1e-4} Eigenfunction of mode 1 for total current (blue curve), runaway current (red curve), and thermal electrons current (green curve) profiles showing the inner layer $\delta_2$ on the runaway current profile and the resistive layer $\delta_1$  on the thermal electrons profile for $\eta=1e-4$, $\Delta'=6.057$ and $c/v_A=1$.}
\end{figure}

\begin{figure}[h]
\includegraphics[width=9 cm, height=6cm]{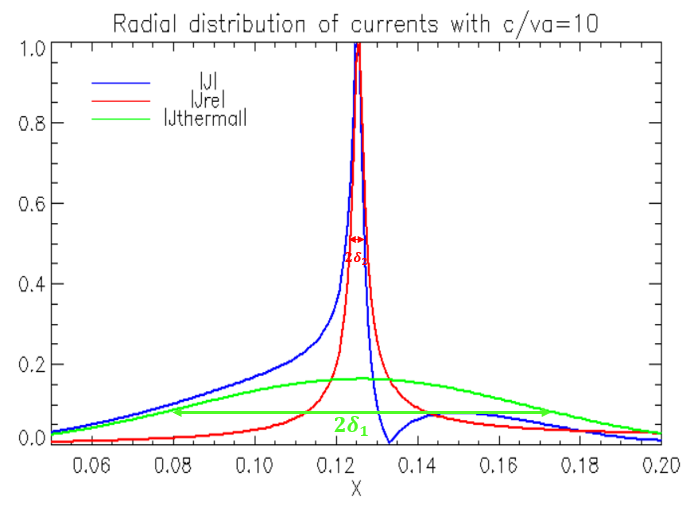}% 
\caption{\label{fig:Delta2_delta1_1e-4_cv10} Eigenfunction of mode 1 for total current (blue curve), runaway current (red curve) and thermal electrons current (green curve) profiles showing the inner layer $\delta_2$ on the runaway current profile and the resistive layer $\delta_1$  on the thermal electrons profile for $\eta=1e-4$, $\Delta'=6.057$ and $c/v_A=10$.}
\end{figure}

Fig. \ref{fig:Delta2_delta1_1e-4} shows the linear eigenfunction for mode 1 of the total current (blue curve), runaway current (red curve), and thermal electrons (green curve) normalized to the maximum of the total current for a case with $\eta=1e-4$,$\Delta'=6.057$, $c/v_A=1$. 
This figure highlights the presence of a resistive layer $\delta_1$ on the thermal current profile and an inner layer $\delta_2$ on the runaway current profile. In particular, it can be observed that the second layer is much narrower than the resistive one and the thermal current profile dominates the runaway current profile. However, as shown in fig. \ref{fig:Delta2_delta1_1e-4_cv10}, which shows the same profiles as in fig. \ref{fig:Delta2_delta1_1e-4} but for $c/v_A=10$, the peak of the runaways is higher than that of the thermal current. This results from the conservation of runaways which in the case of a thinner layer are distributed over a smaller area and reach a maximum that is higher than the peak value of the thermal electrons.\\

\begin{figure}
\includegraphics[width=9 cm, height=6cm]{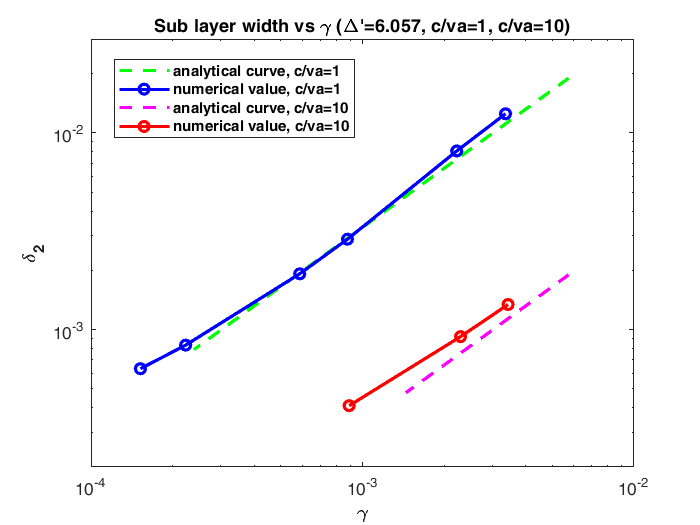}% 
\caption{\label{fig:Sublayer_delta6} Numerically obtained sublayer widths compared with the analytical derivation given in Eq. \ref{delta_2} for different values of $\gamma$ corresponding to $\eta$ in the [1e-4,5e-7] range, $c/v_A=[1,10]$ and $\Delta'=6.057$. The green dashed line corresponds to the analytical curve for $c/v_A=1$ and the magenta dashed line to the analytical curve for $c/v_A=10$, while the blue curve represents the numerical results for $c/v_A=1$ and the red curve for $c/v_A=10$  }
\end{figure}

The dependence of the microlayer $\delta_2$ on the $c/v_A$ ratio is scanned in fig. \ref{fig:Sublayer_delta6} for different values of the resistivity, through the growth rates $\gamma$. In particular, the numerical (blue and red) curves are compared with the anaytical (green and magenta) lines. The blue and green curves correspond to the $c/v_A=1$ case while the red and magenta ones to the $c/v_A=10$ case.  By comparing the two numerical curves we observe  that there is about an order of magnitude difference between the results as it should be according to the definition of $\delta_2$ given in Eq. \ref{delta_2}. For both cases, $c/v_A=1$ and $c/v_A=10$ a good agreement is found between theory and simulations. For the higher ratio, the resistivity interval considered is limited to $\eta=[1e-4,5e-5]$ since below this interval the problem requires a finer resolution which causes the simulations to become computationally unfeasible.

%Apart from the resistivity, the width of the microlayer depends on the $c/v_A$ ratio as can be seen in Eq. \ref{delta_2}. In fig. \ref{fig:Sublayer_delta6}, the numerically determined widths of $\delta_2$ for $c/v_A=[1,10]$ corresponding to the blue curve for the $c/v_A=1$ case and red curve for the $c/v_A=10$ case are compared between them and with the analytical derivation from Eq. \ref{delta_2} represented by the dashed green line for $c/v_A=1$ and magenta dashed line for $c/v_A=10$. The comparison of the two numerical curves for $c/v_A=[1,10]$ shows that there is about an order of magnitude difference between the results as it should be according to the definition of $\delta_2$. For both cases $c/v_A=1$ and $c/v_A=10$ a good agreement is found between the theory and the simulation. For the higher ratio, the resistivity interval considered is limited to $\eta=[1e-4,5e-5]$ since below this interval the problem requires a finer resolution which causes the simulations to become computationally unfeasible.

\begin{figure}
\includegraphics[width=9 cm, height=6cm]{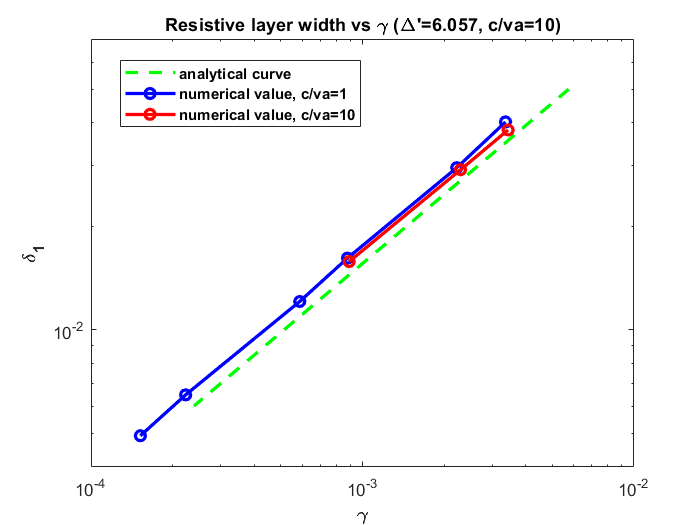}% 
\caption{\label{fig:Resistive_delta6} Numerically obtained resistive layer widths compared with the analytical derivation given in Eq. \ref{delta_1} for different values of $\gamma$ corresponding to $\eta$ in the [1e-4,5e-5] range, $c/v_A=[1,10]$ and $\Delta'=6.057$. The green dashed line corresponds to the analytical curve, while the blue curve represents the numerical results for $c/v_A=1$ and the red curve for $c/v_A=10$.}
\end{figure}

Concerning the resistive layer, fig. \ref{fig:Resistive_delta6}  shows a comparison between the theory (green dashed line) and the numerical results for $c/v_A=[1,10]$ (blue and red curve respectively) and a good agreement is found. Moreover, the curve for $c/v_A=10$ shows a better agreement with the theory with respect to the other curve since a higher ratio makes the simulation more asymptotic.

\begin{figure}
\includegraphics[width=9 cm, height=6cm]{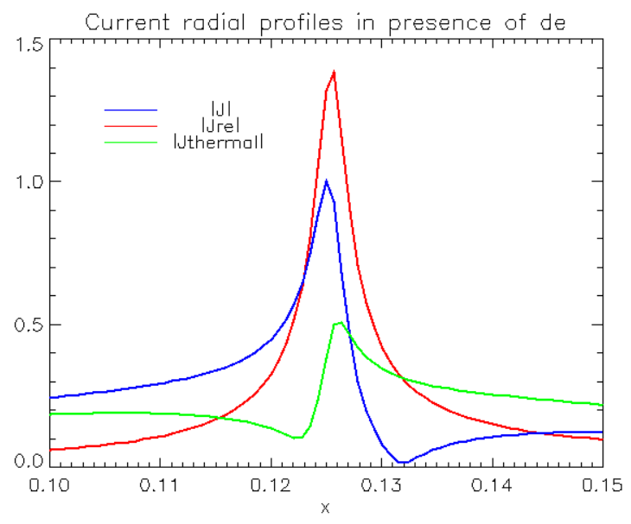}% 
\caption{\label{fig:currents_de} Eigenfunction of mode 1 for total current (blue curve), runaway current (red curve) and thermal electrons current (green curve) profiles for $\eta=1e-4$, $\Delta'=6.057$ and $c/v_A=10$.}
\end{figure}

One question already raised in ref. \cite{Liu2020} concerns the width of the microlayer, which can be comparable with the electron skin depth, $d_e$, and therefore could imply an effect of the electron mass on the  Ohm's law.  In order to investigate the effects of the electron inertia on the system evolution we performed a simulation campaign retaining the terms related to $d_e$ in eq. \ref{psieq}. In particular, a $d_e=0.1$ value was taken into consideration which is close to $d_e=0.017$ taking a post disruptive plasma density of $n_e=1e17 m^{-3}$. While we do not observe any difference when $c/v_A=1$ the presence of $d_e$ in Ohm's law affects the radial distribution of the thermal electrons for $c/v_A=10$ where the runaway current is carrying almost all the plasma current. In this scenario, the thermal electrons are no more characterized by a Gaussian-like distribution as in a purely resistive case, but by a smaller layer as reported in fig. \ref{fig:currents_de}. Comparing fig. \ref{fig:currents_de}, which  shows the eigenfunction of mode 1 for the total current (blue curve), runaway current (red curve), and thermal current (green curve) normalized to the maximum of the total current, with fig.  \ref{fig:Delta2_delta1_1e-4_cv10} the difference between the thermal electron distribution can be appreciated. The presence of the electron skip depth leads the thermal current to become important already during the linear evolution of the island. 

\subsection{\label{asymnonlin} NONLINEAR ANALYSIS}

\begin{figure} [h]
\includegraphics[width=9 cm, height=6cm]{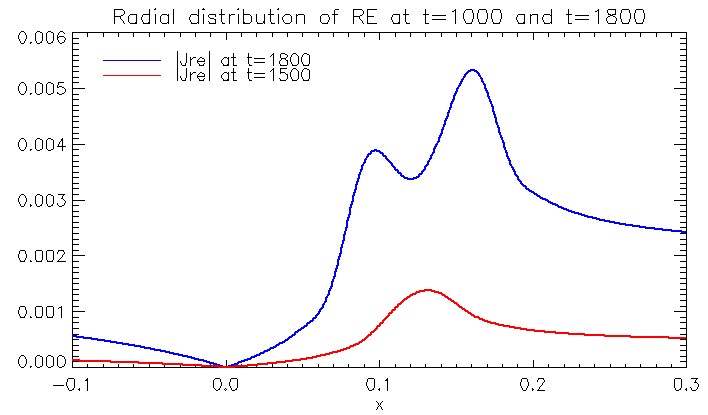}% 
\caption{\label{fig:Microlayer_scission_cv1} Radial profile of the runaway electrons distribution during the linear evolution  (red) at t=1500 and during the non linear evolution at t=1800 of runaways
for $\eta =1e−4$, $c/v_A = 1$ and $\Delta' = 6.057$.}
\end{figure}

\begin{figure} [h]
\includegraphics[width=9 cm, height=6cm]{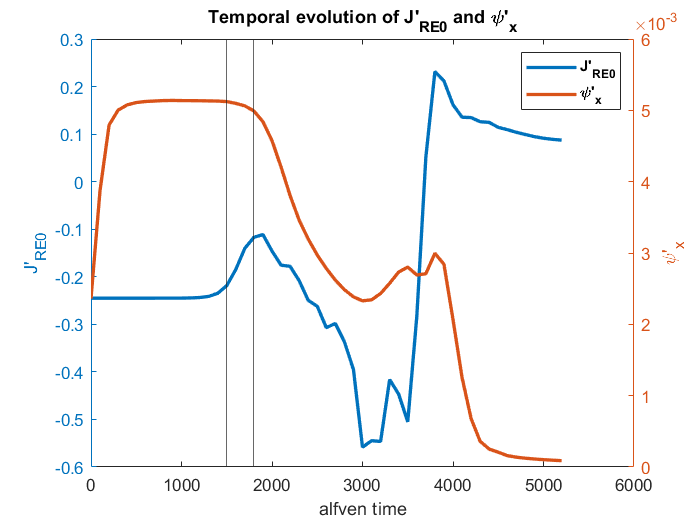}% 
\caption{\label{fig:PSI_JRE_CV1} Temporal evolution of $\psi_X$ (rust) and $J'_{RE0}$ (blue)  for $\eta=1e-4$, $c/v_A=1$ and $\Delta'=6.057$.  }
\end{figure}

With respect to the linear regime, where a smaller integration domain does not affect  the evolution of the system,  in the nonlinear regime the island width reaches dimensions of the order of the domain radial extension. As a consequence, in order to avoid border effects, an extension of $Lx=3\pi$ was chosen with $\eta=1e-4$ and $nx=4800$. This setup enables us to have enough spatial resolution in the reconnection region even with a larger domain.\\
During the non-linear evolution of the magnetic reconnection process in the presence of a runaway current, the distribution of the RE population undergoes significant changes. Specifically, when entering the nonlinear phase the single microlayer observed on the RE profile splits into multiple local peaks as shown in fig. \ref{fig:Microlayer_scission_cv1} where the eigenfunction of mode 1 associated with RE radial profiles at t=1500 (red) and $t=1800$ (blue) are depicted. At $t=1500$ the runaway electron evolution is at the end of the linear regime and at $t=1800$ it is in the nonlinear regime as shown by the vertical lines in fig. \ref{fig:PSI_JRE_CV1}. Here we show the temporal evolution of $\psi$ at the X-point (rust), which in the linear phase is directly  linked to the island evolution,  and the temporal evolution of the derivative of the eigenfunction of mode 0 representing the equilibrium runaway current at the rational surface, $J'_{RE0}$ (blue) for  $\eta=1e-4$, $c/v_A=1$ and $\Delta'=6.057$. Moreover, it was observed that with a higher value of $c/v_A$  the runaways become nonlinear at an earlier stage. Indeed, with $c/v_A=10$, the runaways become nonlinear already during the linear evolution of the island. It was found that the transition from the linear to the nonlinear regime for runaways is governed by the widths of the inner layer $\delta_2$ and the island. In particular, when the island width becomes larger than the inner layer the runaways become non linear and as the inner layer gets smaller with increasing $c/v_A$, the RE become non linear earlier in the case with $c/v_A=10$ with respect to the case where $c/v_A=1$.

%\begin{figure} [h]
%\includegraphics[width=9 cm, height=6cm]{JRE0_PSI_X_CV10.png}% 
%\caption{\label{fig:PSI_JRE_CV10} Temporal evolution of $\psi_X$ (rust) and $J'_{RE0}$ (blue)  for %$\eta=1e-4$, $c/v_A=10$ and $\Delta'=6.057$.  }
%\end{figure}

\begin{figure*}
\includegraphics[width=\textwidth,height=7cm]{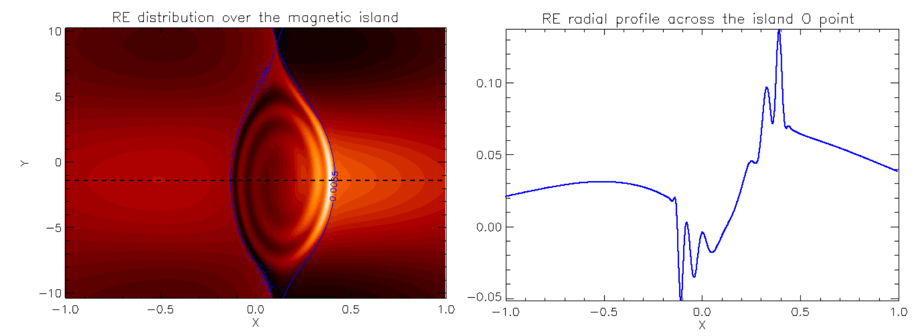}% 
\caption{\label{fig:Spiral} Runaway electrons distribution over the magnetic island during the non linear evolution of the runways (left) and runaway electron profile at t=2500 (right) for $\eta=1e-4$, $c/v_A=1$ and $\Delta'=6.057$ at Y=-5.  }
\end{figure*}

In contrast to the linear phase where the RE concentrates on the X-point of the island, during the nonlinear phase, the RE starts firstly to distribute over the separatrices of the island, and then, advancing more and more in the nonlinear phase, over multiple peaks. The RE's redistribution during its nonlinear evolution is significantly impacted by the combination of two phenomena which are the island growth and its rotation, resulting in the formation of a spiral-like structure as represented in fig. \ref{fig:Spiral} where the left figure depicts the RE distribution over a magnetic island and the right figure shows the RE radial distribution along the spiral for the $\eta=1e-4$, $c/v_A=1$ and $\Delta'=6.057$ case across the island O point and $t=2500$. As the island grows and the runaways distribute over the separatrices of the island, these are influenced by the island's rotation along the poloidal direction to form a spiral-like structure. In the case where $c/v_A=10$ the island width becomes larger than $\delta_2$ already during the linear phase of the island evolution causing the runaways to become nonlinear which leads to the formation of the spiral at an earlier stage with respect to the $c/v_A=1$  case.\\

\begin{figure*}
\includegraphics[width=\textwidth,height=7cm]{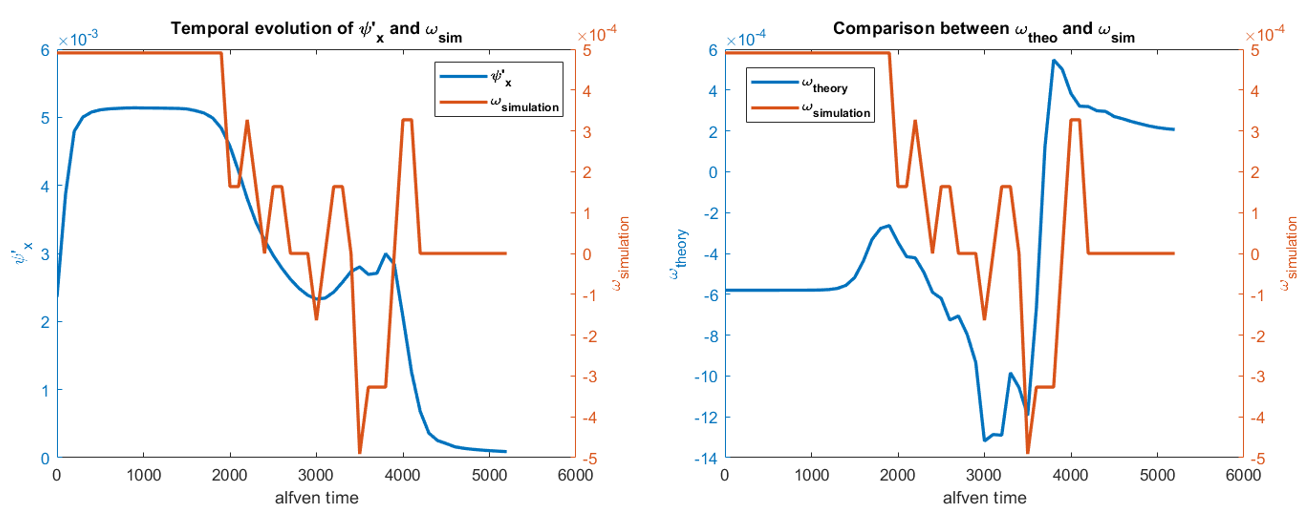}% 
\caption{\label{fig:Rotationfrequency_cv1} Temporal evolution of $\omega_{simulation}$ (rust) and $\psi'_x$ (blu) at left and temporal evolution of $\omega_{theory}$ (blue) and $\omega_{simualtion}$ (rust) at right for $\eta=1e-4$, $c/v_A=1$ and $\Delta'=6.057$.  }  
\end{figure*}

The spiral-like structure ceases to exist once the island rotation tends toward zero. Indeed, as it can be observed in the left panel of fig. \ref{fig:Rotationfrequency_cv1} which compares the evolution of the island by the mean of $\psi'_x$  (blue curve) and its rotation frequency (rust curve) when the island enters the nonlinear regime after $t=2000$ the island rotation tends towards zero. Once the island saturates, approximately around $4000\tau_A$, also the rotation frequency goes to zero. Moreover, the rotation frequency depends strictly on $J'_{RE0}$, so it agrees with the eq. \ref{Gr}, only during the linear phase of the runaway current evolution, as it can be observed in the right panel of fig. \ref{fig:Rotationfrequency_cv1}. Here, the analytical value,  $\omega_{theory}$ (blue curve),  computed using eq. \ref{Gr} with $J'_{RE0}$ given by the simulation, is compared with the numerical rotation frequency,  $\omega_{simulation}$ (rust curve), for $\eta=1e-4$, $c/v_A=10$ and $\Delta'=6.057$. We can clearly see that the agreement between the two curves stops when the equilibrium runaway current is perturbed. As a consequence, computing $\omega_{theory}$ through eq. \ref{Gr} leads to a non-zero rotation frequency at saturation whereas in the simulation this goes to zero. 

\begin{figure*}
\includegraphics[width=\textwidth,height=7cm]{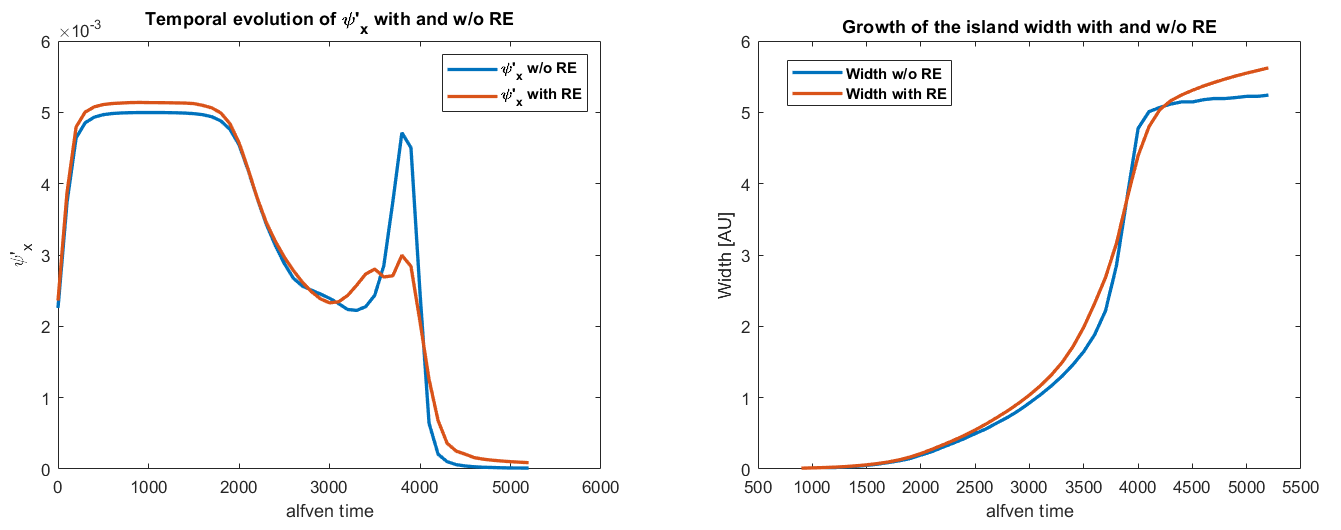}% 
\caption{\label{fig:Saturation_delta6} Temporal evolution of $\psi'_X$ w/o (blue curve) and with RE (rust curve)   at left panel  and temporal evolution of island area w/o (blue curve) and with RE (rust curve) for at the right panel for $\eta=1e-4$, $c/v_A=1$ and $\Delta'=6.057$.  }  
\end{figure*}

\begin{figure*}
\includegraphics[width=\textwidth,height=7cm]{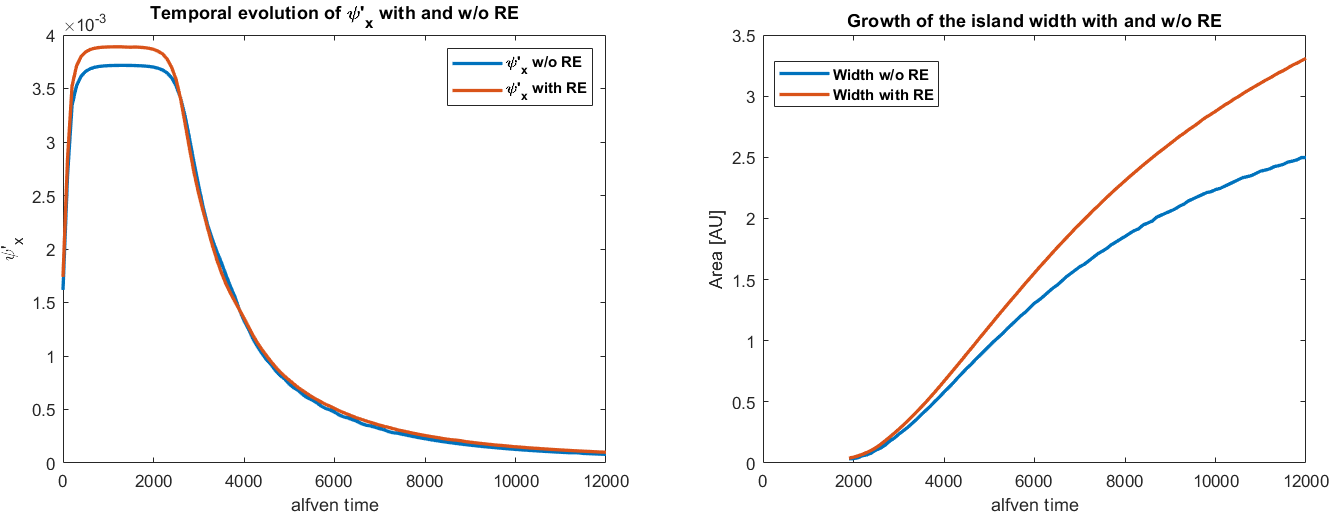}% 
\caption{\label{fig:Saturation_delta3} Temporal evolution of $\psi'_X$ w/o (blue curve) and with RE (rust curve)   at left panel  and temporal evolution of island area w/o (blue curve) and with RE (rust curve) at the right panel for $\eta=1e-4$, $c/v_A=1$ and $\Delta'=3$.  }  
\end{figure*}

Concerning the island width at saturation, it was found that with $\Delta'=6$  the presence of a runaway current does not lead to a $50\%$ higher width of the saturated island with respect to the width in the absence of runaways as demonstrated in \cite{Helander2007} and shown in fig. \ref{fig:island}. Indeed, by comparing the simulations with and without runaways it was observed that when $\Delta'>3$ the saturated island width in the presence of RE was comparable to the dimension of the island in the absence of RE. This is caused by a different nonlinear evolution of the island as it can also be noticed by looking at the temporal evolution of $\psi'_x$ in fig. \ref{fig:PSI_JRE_CV1} which tends to increase in the nonlinear phase. Without RE, the growth of $\psi'_X$, which approximates well the island growth even in the nonlinear phase,  is more pronounced as can be noticed in the left panel in fig. \ref{fig:Saturation_delta6}, which depicts the temporal evolution of $\psi'_X$ w/o (blue curve) and with RE (rust curve). This burst in the island growth only takes place in the absence of RE, compensating for the $50\%$ larger island seen in the presence of RE. Thus this behavior leads to a magnetic island at saturation similar in size, as can be seen in the right panel of fig. \ref{fig:Saturation_delta6}.   For the sake of completeness, we also show 
the same plots for the case $\Delta'=3$ in fig. \ref{fig:Saturation_delta3}.  In this case, the nonlinear  evolution of $\psi'_X$ is characteristic of the small $\Delta'$ regime and leads to a magnetic island width at saturation approximately $50\%$ higher in presence of RE compared to the case without RE as predicted by the theory in \cite{Helander2007}. 

\section{\label{sec:level6} Conclusions}
In this paper we have addressed the problem of the stability of a post-disruption plasma in a realistic asymmetric configuration, characterised by a mismatch between the current peak and the resonant surface.
 The runaway fluid equation is coupled with the MHD equations through a current coupling scheme and all the equilibrium plasma current is assumed to be carried by the runaways.\\ 
We find that, while, as in the symmetric case \cite{Helander2007}, the runaways do not alter the linear growth of the island, they lead to a rotation of the island in the poloidal direction consistently with the analytical results shown in \cite{Liu2020}. An additional feature seen in the asymmetric simulations is the microlayer on the runaway current profile much smaller than the resistive layer. While the resistive layer controls the transition of the island from the linear to the nonlinear stage, the microlayer width causes the runaways to become nonlinear once the island size becomes larger than the microlayer width. This transition of the  runaways to the nonlinear phase is accompanied by the generation of a spiral-like structure inside the island changing drastically the distribution of runaways with respect to the symmetric case. In addition, since the microlayer widths are comparable in size with the electron skin depths measured in a post-disruptive scenario we studied the effect of the presence 
 of $d_e$ during the linear evolution of the island and find that the thermal electron distribution is no more characterized by the presence of a resistive layer, but a much narrower layer. As said earlier, the resistive layer governs the island transition to the nonlinear phase so it can be assumed that this change in the thermal electrons distribution might have an influence on the island growth as well. However, a nonlinear analysis of a RE-driven magnetic reconnection in the presence of $d_e$ was not possible due to numerical noise rising  when approaching the nonlinear stage of the island.\\
The nonlinear analysis in the resistive regime shows that the frequency does not follow the evolution of the equilibrium runaway current once this becomes nonlinear and tends to zero when the island goes toward saturation.
Finally, we find that as far as we are in the small $\Delta'$ regime the island width at saturation is $50\%$ bigger than the corresponding island without runaways, consistently with the theory \cite{Helander2007}.
While, increasing $\Delta'$,  the magnetic island saturation width in the presence of RE becomes more similar in size to the width in the absence of RE.
In this case the rapid burst in the nonlinear island growth in the absence of RE compensates the $50\%$ higher saturation width with RE.\\
This study confirms the importance of investigating the stability of a post disruptive plasma in the presence of a RE current. Among other effects, the presence of more than one helicity in the initial perturbation can lead to 
plasma stochastization, connecting strictly this work with the study of the benign runaway termination in presence of stochastic magnetic field \cite{Bandaru2021}. 
 On top of that, a $50\%$ higher width at saturation of the magnetic island may have a significant effect on the evolution of the magnetic stochasticity preventing the RE population growth. These aspects of the RE driven magnetic reconnection will be addressed in a future study. 

\section*{ Acknowledgement}

The numerical simulations were performed using the EUROfusion high performance computer Marconi Fusion  hosted at CINECA (Project No. FUA36-FKMR2)

\section*{ Data Availability}
The data that support the findings of this study are available from the corresponding author
upon reasonable request.

\newpage
\nocite{*}
\bibliography{aipsamp}% Produces the bibliography via BibTeX.

\end{document}